\documentclass[a4paper,fleqn]{cas-dc}
\usepackage[version=4]{mhchem}
\usepackage{amsmath}
\usepackage{hyperref}
\usepackage[authoryear,longnamesfirst]{natbib}
\usepackage{lineno}
\usepackage{adjustbox}
\def\tsc#1{\csdef{#1}{\textsc{\lowercase{#1}}\xspace}}
\tsc{WGM}
\tsc{QE}
\tsc{EP}
\tsc{PMS}
\tsc{BEC}
\tsc{DE}
\begin{document}
\let\WriteBookmarks\relax
\def\floatpagepagefraction{1}
\def\textpagefraction{.001}
\shorttitle{Ethylene oxide and acetaldehyde towards G358.93--0.03 MM1}
\shortauthors{Manna \& Pal}
%\begin{frontmatter}
\title [mode = title]{First detection of ethylene oxide and acetaldehyde in hot core G358.93--0.03 MM1: Tracing prebiotic oxygen chemistry}  
	
\author[1]{Arijit Manna}
\ead{amanna.astro@gmail.com, tel: +918777749445}
\address[1]{Department of Physics and Astronomy, Midnapore City College, Paschim Medinipur, West Bengal  721129, India}
\cormark[1]
\author[1]{Sabyasachi Pal}
%\cormark[1]
	
\begin{abstract}
Ethylene oxide (c-\ce{C2H4O}) and its isomer, acetaldehyde (\ce{CH3CHO}), are important complex organic molecules owing to their potential role in the formation of amino acids (R--CH(\ce{NH2})-COOH) in ISM. The detection of c-\ce{C2H4O} in hot molecular cores suggests that the possible existence of larger ring-shaped molecules containing more than three carbon atoms, such as furan (c-\ce{C4H4O}), which shares structural similarities with ribose (\ce{C5H10O5}), the sugar component of DNA. In this study, we report the first detection of the rotational emission lines of c-\ce{C2H4O} and \ce{CH3CHO} towards the hot molecular core G358.93--0.03 MM1, based on observations from the Atacama Large Millimeter/Submillimeter Array (ALMA) in band 7. The fractional abundances of c-\ce{C2H4O} and \ce{CH3CHO} relative to \ce{H2} are $(2.1\pm0.2)\times10^{-9}$ and $(7.1\pm0.9)\times10^{-9}$, respectively. The column density ratio between \ce{CH3CHO} and c-\ce{C2H4O} is $3.4\pm0.7$. A Pearson correlation heat map reveals strong positive correlations ($r$ $>$ 0.5) between the abundances and excitation temperatures of c-\ce{C2H4O} and \ce{CH3CHO}, suggesting a possible chemical connection between those two molecules. To investigate this further, we conducted a two-phase warm-up chemical model using the gas-grain chemical code UCLCHEM. A comparison between our derived abundances and the predictions from our chemical model and existence model demonstrates good agreement within factors of 0.73 and 0.74, respectively. We propose that c-\ce{C2H4O} may form in G358.93--0.03 MM1 via the grain surface reaction between \ce{C2H4} and O, but \ce{CH3CHO} may be produced through the surface reaction between \ce{CH3} and HCO.
\end{abstract}
	
\maketitle
	
\section{Introduction}
\label{introduction}
Understanding the chemical formation pathways and organic compositions of dense molecular clouds is crucial for unravelling the astrochemical evolution of various celestial bodies, including asteroids, comets, and protoplanetary disks \citep{eh00}. Heterocycles, a class of complex chemical compounds, contain heavier atoms in their ring structures in addition to carbon. Among these, the study of oxygen (O)-bearing species in the interstellar medium (ISM) is particularly significant because of their potential association with the origins of life \citep{occ14}. Ribose (\ce{C5H10O5}) is a key molecule linked to the molecular structure of DNA. A simpler heterocyclic molecule in comparison to \ce{C5H10O5} is ethylene oxide (c-\ce{C2H4O}), also known as oxirane. 
		
\begin{figure}
\centering
\includegraphics[width=0.5\textwidth]{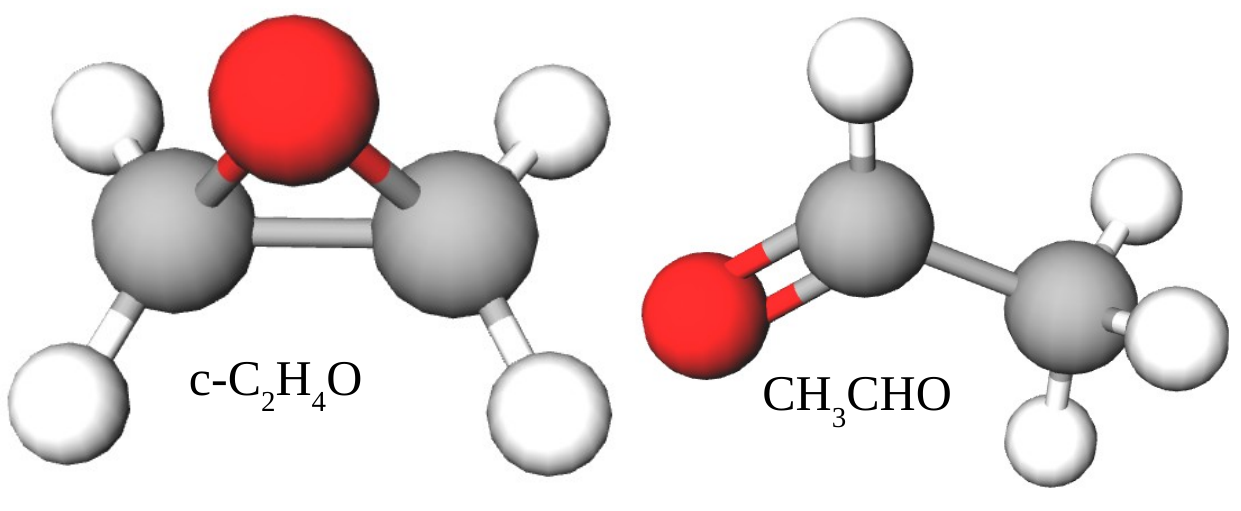}
\caption{Molecular structures of c-\ce{C2H4O} and \ce{CH3CHO}. The grey, white, and red spheres represent the C, H, and O atoms, respectively.}
\label{fig:molecule}
\end{figure}

In the ISM, c-\ce{C2H4O} is the simplest ring-shaped epoxide organic molecule in which an oxygen atom is bonded to two carbon atoms. Due to its two pairs of identical interchangeable hydrogen nuclei, c-\ce{C2H4O} exhibits two distinct symmetry states, referred to as ortho (symmetric) and para (anti-symmetric). These states correspond to the energy levels characterized by the quantum numbers \( K_a K_c \), with ee/oo representing the ortho state and eo/oe representing the para state \citep{nu98}. Their respective spin weights are 10 and 6. The rotational spectrum of c-\ce{C2H4O} is purely b-type (\(\mu_b = 1.88\) D) \citep{cu51}. The three-dimensional molecular structure of c-\ce{C2H4O} is shown in Figure~\ref{fig:molecule}, which is computed using MolView \citep{ber15}. Previous studies have indicated that c-\ce{C2H4O} plays a crucial role in the synthesis of amino acids and other prebiotic compounds in the ISM environment \citep{cl02, mil93}. The rotational emission lines of c-\ce{C2H4O} were first detected in the ISM toward the Sgr B2 (N) molecular cloud \citep{dic97}. Subsequently, \cite{nu98} and \cite{ik01} conducted surveys to study the emission lines of c-\ce{C2H4O} in 20 hot cores and two dark clouds. After spectral analysis, \cite{nu98} and \cite{ik01} found that the abundance of c-\ce{C2H4O} in hot cores is on the order of $\sim$10$^{-10}$. However, this molecule was not detected in the dark clouds of TMC-1 and TMC-1 (\ce{NH3}). In addition to hot cores, the rotational emission lines of c-\ce{C2H4O} have also been detected in prestellar cores \citep{bac19}, warm Galactic center sources \citep{re08}, and hot corino IRAS 16293--2422 B \citep{ly17}. Moreover, mono-deuterated c-\ce{C2H3DO} emission lines were observed toward IRAS 16293-2422 B \citep{mu23}. Earlier, \cite{dic97} proposed that c-\ce{C2H4O} forms in the gas-phase, where ethanol (\ce{C2H5OH}) acts as one of the reactants: \\\\ \ce{CH3}$^{+}$ + \ce{C2H5OH} $\longrightarrow$ \ce{C2H5O}$^{+}$ + \ce{CH4}~~~~~~~~~~~~(1) \\\\
\ce{C2H5O}$^{+}$ + e$^{-}$ $\longrightarrow$ \ce{C2H4O} + H~~~~~~~~~~~~~~~~~~~~~~~~~(2) \\\\ 
To verify reactions (1) and (2), \cite{ik01} conducted chemical simulations of c-\ce{C2H4O} using pure gas-phase reactions at different gas temperatures and densities. Their simulations revealed that the modelled gas-phase abundance of c-\ce{C2H4O} decreases with increasing gas kinetic temperature. Consequently, \cite{ik01} also concluded that grain surface reactions are crucial for synthesising of c-\ce{C2H4O}.
		
Acetaldehyde (\ce{CH3CHO}) is a non-cyclic isomer of c-\ce{C2H4O}, serving as an essential evolutionary tracer for various complex O-bearing molecules in the ISM \citep{her09}. Previous quantum chemical studies showed that \ce{CH3CHO} exhibits both \textit{a}- and \textit{b}-type transitions ($\mu_{a} = 2.423$ D, $\mu_{b} = 1.266$ D) \citep{bau76}. Additionally, the internal rotation of its methyl (\ce{CH3}) group results in two non-interacting torsional substates, denoted $A$ and $E$, with their ground-state energies separated by 0.1 K. The molecular diagram of \ce{CH3CHO} is shown in Figure~\ref{fig:molecule}. The emission lines of \ce{CH3CHO} were first detected in the cold dust clouds TMC-1 and L134N, as well as in the massive star-forming region Sgr B2 \citep{mat85}. Subsequently, \cite{nu98} and \cite{ik01} investigated the emission lines of \ce{CH3CHO} in more than 20 hot cores and found that their abundances are of the order of $\sim$10$^{-9}$. The rotational emission lines of \ce{CH3CHO} and its isotopologues \ce{CH2DCOH}, \ce{CH3COD}, and \ce{CHD2CHO} were detected in the hot core object IRAS 16293--2422 B using the ALMA-PILS survey \citep{jo18, cou19, fe23}. Earlier, \cite{nu98} and \cite{ik01} found that \ce{CH3CHO} is spatially correlated with methanol (\ce{CH3OH}). \cite{re08} reported that \ce{CH3OH}/\ce{CH3CHO} ratio is approximately $\sim 10$ in dark clouds and $\sim 100$ in hot cores.  
		
\begin{figure}
\centering
\includegraphics[width=0.4\textwidth]{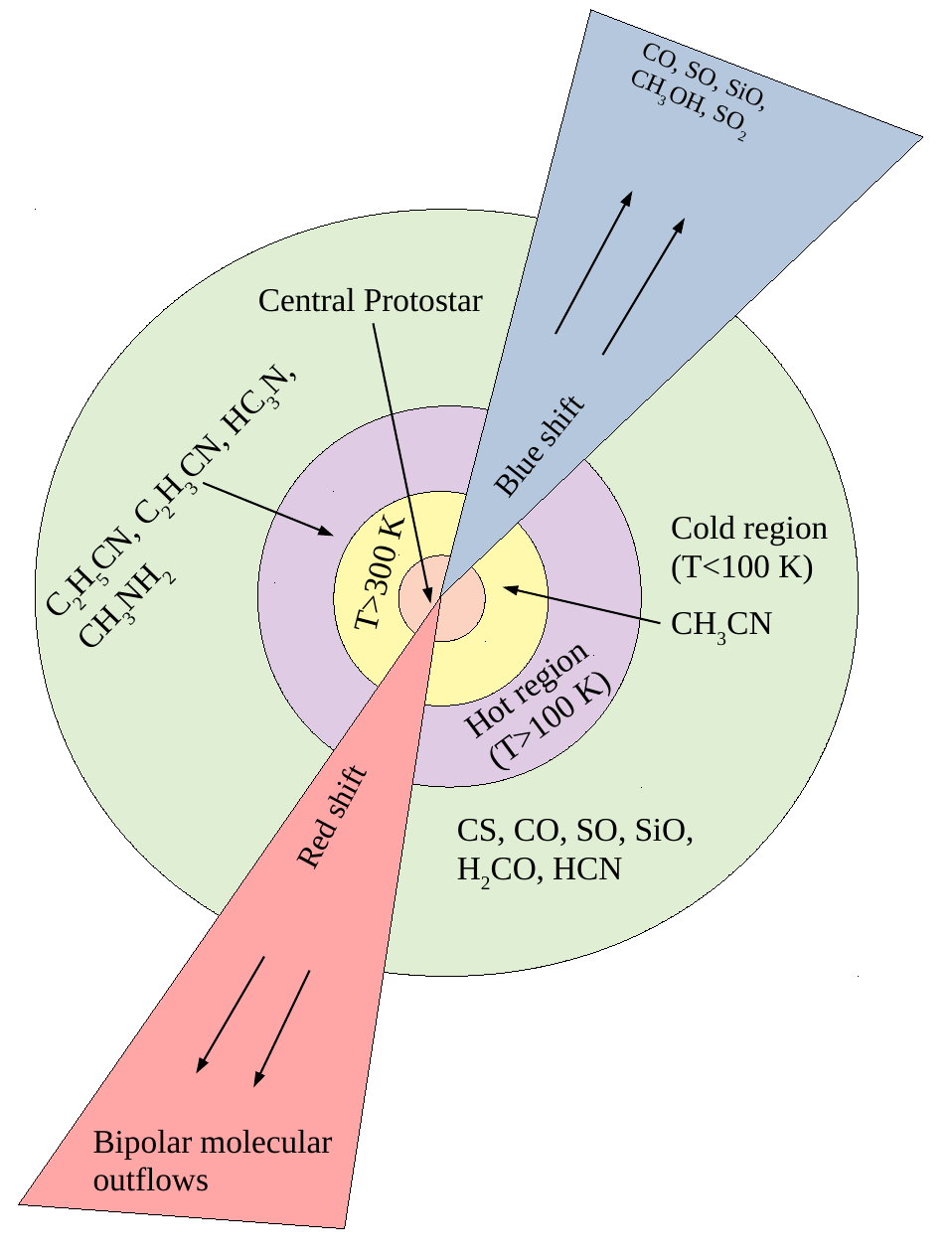}
\caption{Schematic diagram and molecular distribution of the hot molecular core.}
\label{fig:hotcore}
\end{figure}
		
\begin{figure*}
\centering
\includegraphics[width=1.0\textwidth]{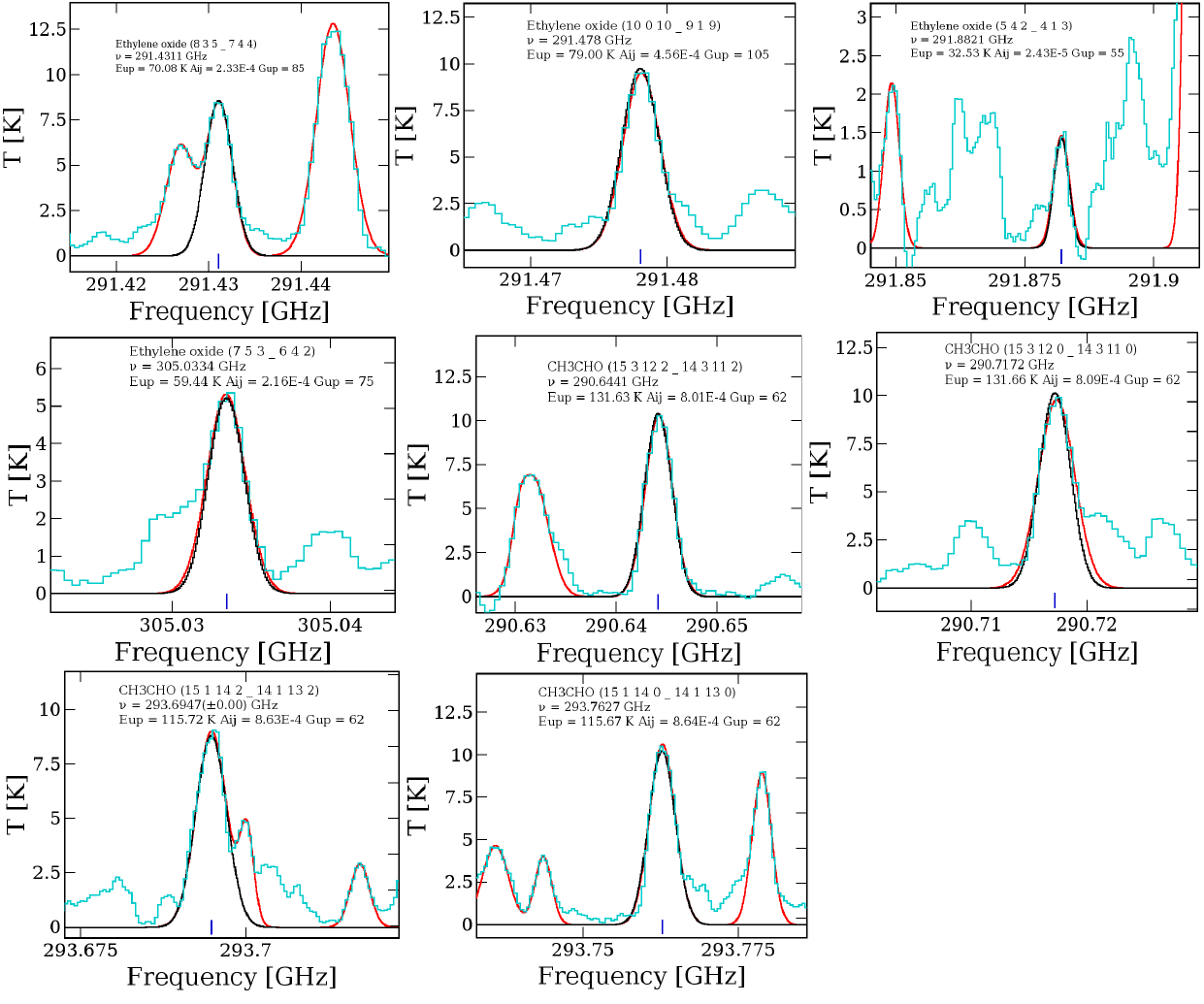}
\caption{Molecular emission lines of c-\ce{C2H4O} and \ce{CH3CHO} towards G358.93--0.03 MM1. The blue lines represent the observed spectra, whereas the black lines correspond to the LTE model spectra of c-\ce{C2H4O} and \ce{CH3CHO}. The red lines depict the LTE model spectra of other molecules identified in the region. The quantum numbers for each transition of both molecules are given in the form $J_{K_a,K_c} \rightarrow J'_{K'_a,K'_c}$, where $J$ is the total rotational angular momentum, and $K_a$, $K_c$ are its projections along the principal axes of molecules.}
\label{fig:spectra}
\end{figure*}
		
It is widely recognized that most stars, particularly high-mass stars ($>8 M_{\odot}$), develop into dense clusters \citep{car00, lada03, riv13}. Hot molecular cores (HMCs) are associated with early stages of high-mass star formation regions \citep{her09}. While there is still much to learn about the birth of high-mass stars, the following evolutionary model has emerged from recent observational studies: Infrared dark clouds $\rightarrow$ HMCs $\rightarrow$ hyper/ultra-compact H\,\textsc{ii} regions $\rightarrow$ The H\,\textsc{ii} region around high-mass stars that are ionizing \citep{beu07}. HMCs represent a relatively compact stage in the high-mass star formation process ($\leq0.1$ pc), typically characterized by high gas densities $\geq10^{7}$ cm$^{-3}$ and temperatures of $\geq100$ K \citep{kur00}. The sublimation of \ce{H2O} and organic-rich ice mantles results in a wide range of complex organic molecules (COMs), including \ce{C2H5OH}, \ce{CH3CN}, \ce{C2H5CN}, \ce{CH3OH}, and \ce{CH3OCHO}, which have been detected in numerous HMCs \citep{her09, bel13, manet24a}. Previous warm-up chemical models suggest that HMCs are ideal environments for searching for the simplest amino acid glycine (\ce{NH2CH2COOH}), and its various precursors \citep{gar08, gar13, suz18, zh24}. In addition, the emission lines of \ce{NH2CH2COOH} precursor molecules such as methanimine (\ce{CH2NH}), cyanamide (\ce{NH2CN}), methylamine (\ce{CH3NH2}), and amino acetonitrile (\ce{NH2CH2CN}) have been detected toward several hot core candidates \citep{man22a, man22b, suz23, man24a, man24b, man24c}. A schematic diagram and the molecular distribution in HMC, based on the observations of \cite{shi21}, \cite{san22}, and \cite{suz23}, are shown in Figure~\ref{fig:hotcore}. 
		
The massive star-forming region G358.93--0.03 is situated at a distance of 6.75$\,{}\,^{\,+0.37}_{\,-0.68}$\, kpc, with an estimated gas mass of 167$\pm$12 \textup{M}$_{\odot}$ and a luminosity of 7.7$\times$10$^{3}$ \textup{L}$_{\odot}$ \citep{re14, bro19}. This region hosts eight known dust continuum sources, labelled as G358.93--0.03 MM1 to G358.93--0.03 MM8.  Within this star-forming region, G358.93--0.03 MM1 and G358.93--0.03 MM3 are both HMC candidates \citep{bro19}. A variety of complex molecular species have been detected toward G358.93--0.03 MM1, such as cyanamide (\ce{NH2CN}) \citep{man23}, glycolaldehyde (\ce{CH2OHCHO}) \citep{manet23}, ethylene glycol ((CH$_2$OH)$_2$) \citep{manet24}, formamide (\ce{NH2CHO}), isocyanic acid (HNCO) \citep{man24d}, and methylamine (\ce{CH3NH2}) \citep{man24a}. The presence of such a chemically diverse inventory highlights that G358.93--0.03 MM1 is a promising site for investigating different complex biomolecules.
		
This paper is organized as follows: Section~\ref{obs} outlines the ALMA observations and data reduction methods. The identification of c-\ce{C2H4O} and \ce{CH3CHO} and their chemical modelling are shown in Section~\ref{res}. Finally, Sections~\ref{dis} and \ref{con} present the discussions and conclusions of the study.
	
\begin{table*}
\centering
		%\scriptsize 
\caption{Spectral line parameters of c-\ce{C2H4O} and \ce{CH3CHO} towards G358.93--0.03 MM1.}
\begin{adjustbox}{width=0.92\textwidth}
\begin{tabular}{cccccccccccc}
\hline 
Molecule&Rest frequency &Transition & $E_{u}$ & $A_{ij}$ &g$_{up}$&$S\mu^{2}$&$\int{T_{mb}}dV$ &FWHM &Optical depth\\
&(GHz) & &(K)&(s$^{-1}$) & &(Debye$^{2}$)&K km s$^{-1}$ &(km s$^{-1}$) &($\tau$)\\
\hline
c-\ce{C2H4O}&291.431&8(3,5)--7(4,4)&70.08&2.33$\times$10$^{-4}$&85&68.70&27.02$\pm$3.52 &3.52$\pm$0.25&0.15\\ 
			&291.478&10(0,10)--9(1,9)&79.01&4.56$\times$10$^{-4}$&105&166.09&28.22$\pm$1.56 &3.51$\pm$0.32&0.25\\
			&291.882&5(4,2)--4(1,3)&32.53&2.43$\times$10$^{-5}$&55&4.52&5.23$\pm$0.23 &3.51$\pm$0.28&0.23\\
			&305.033&7(5,3)--6(4,2)&59.44&2.16$\times$10$^{-4}$&75&48.96&14.62$\pm$0.82 &3.51$\pm$0.28&0.20\\		
\hline
\ce{CH3CHO}&290.644&15(3,12)--14(3,11)$E$&131.63&8.01$\times$10$^{-4}$&62&180.52&34.46$\pm$2.32 &3.56$\pm$0.18&0.16 \\
		   &290.717&15(3,12)--14(3,11)$A$&131.66&8.09$\times$10$^{-4}$&62&182.16&32.42$\pm$1.68 &3.55$\pm$0.12&0.12 \\
		   &293.694&15(1,14)--14(1,13)$E$&115.72&8.63$\times$10$^{-4}$&62&188.59&39.88$\pm$2.21 &3.55$\pm$0.26&0.15\\
		   &293.762&15(1,14)--14(1,13)$A$&115.67&8.64$\times$10$^{-4}$&62&188.51&48.62$\pm$3.23 &3.57$\pm$0.32&0.17\\	
\hline
\end{tabular}	
\end{adjustbox}
\label{tab:MOLECULAR DATA}
\end{table*}
		
\begin{figure*}
\centering
\includegraphics[width=0.98\textwidth]{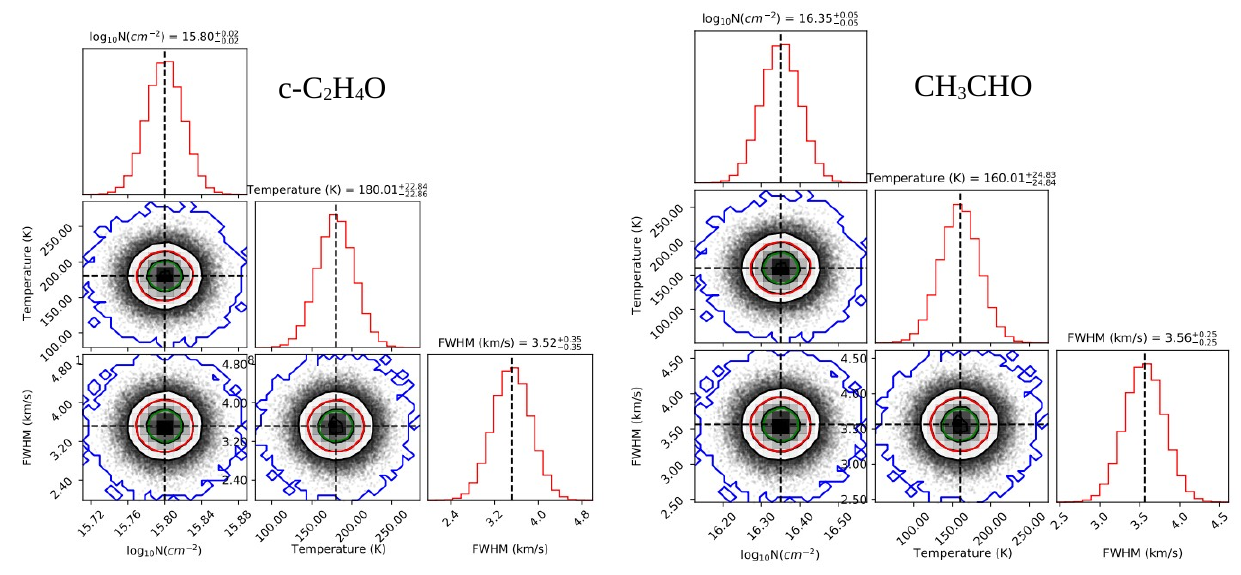}
\caption{Corner plots illustrating the covariances among the posterior probability distributions of the column density (log$_{10}$($N$) in cm$^{-2}$), excitation temperature (in K), and FWHM (in km s$^{-1}$) for c-\ce{C2H4O} and \ce{CH3CHO}.}
\label{fig:corner}
\end{figure*}
		
\begin{figure*}
\centering
\includegraphics[width=1.0\textwidth]{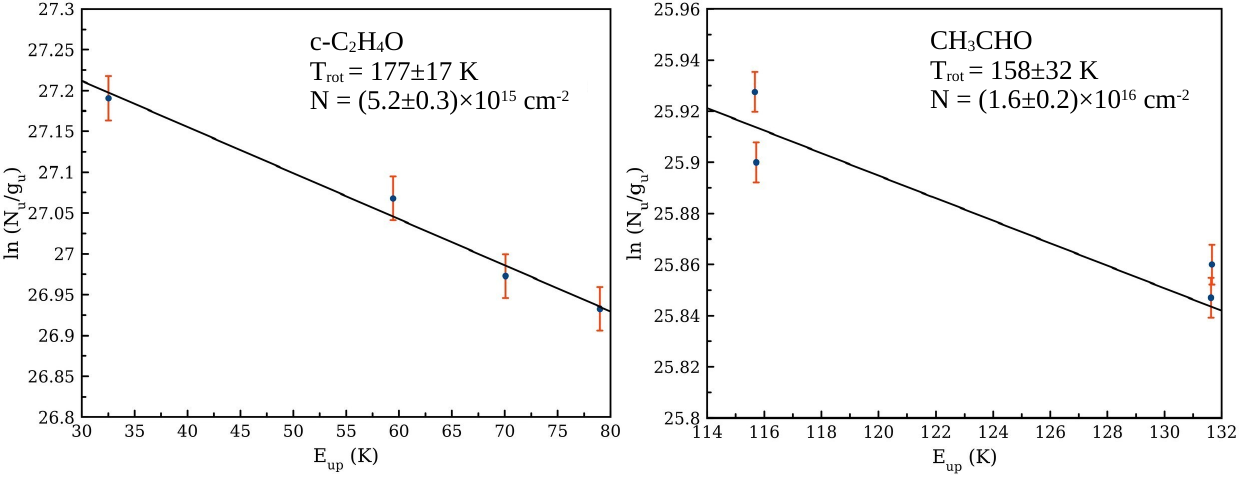}
\caption{Rotational diagram of c-\ce{C2H4O} and \ce{CH3CHO} for the estimation of the total column density ($N_{T}$ in cm$^{-2}$) and rotational temperature ($T_{rot}$ in K). The blue dots are the data points, and the red vertical lines are the error bars. The black lines are the best-fitted straight line.}
\label{fig:rotd}
\end{figure*}
		
\begin{figure*}
\centering
\includegraphics[width=1.0\textwidth]{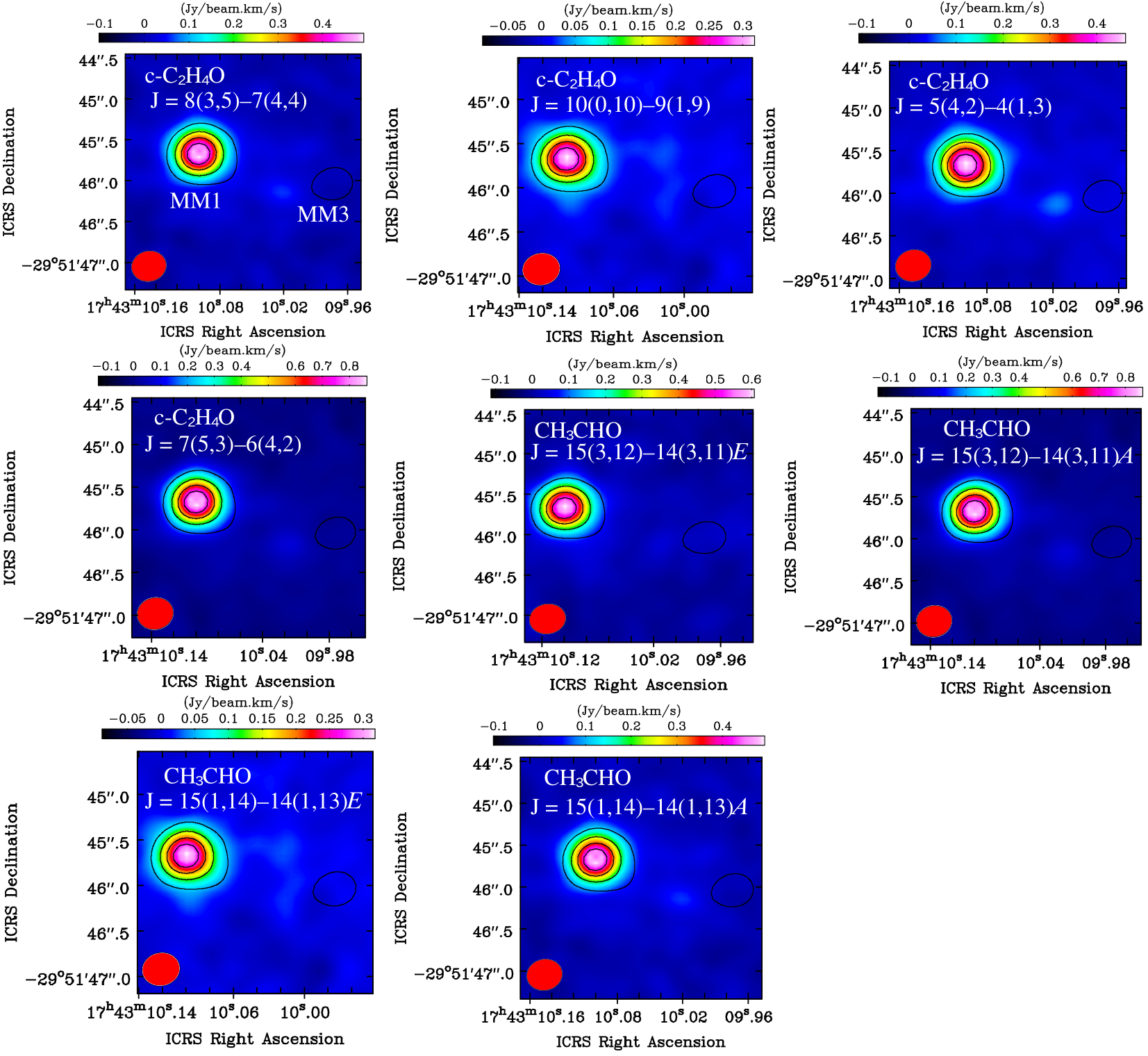}
\caption{Integrated emission (moment-zero) maps of c-\ce{C2H4O} and \ce{CH3CHO} toward G358.93--0.03 MM1, overlaid with 988 $\mu$m continuum emission map (black contours). The contour levels are set at 20\%, 40\%, 60\%, and 80\% of the peak flux. The red circles indicate the synthesized beams corresponding to the integrated emission maps.}
\label{fig:intmap}
\end{figure*}

\section{Observations and data reduction}
\label{obs}
The observations of the high-mass star-forming region G358.93--0.03 were carried out using the Atacama Large Millimeter/submillimeter Array (ALMA) band 7 receivers, as part of a study on high-mass protostellar accretion outbursts (ID: 2019.1.00768.S; PI: Crystal Brogan). The observation was conducted on October 11, 2019, and it took a total on-source integration time of 756 seconds. A total of 47 antennas were used, spanning the baselines from 14 m to 2517 m. Flux and bandpass calibrations were performed using J1550+0527, whereas J1744-3116 served as the phase calibrator. The observed frequency coverage included four spectral windows: 290.51--292.39 GHz, 292.49--294.37 GHz, 302.62--304.49 GHz, and 304.14--306.01 GHz. The corresponding spectral and velocity resolutions were 1128.91 kHz and 0.96 km s$^{-1}$, respectively.
		
Data calibration and imaging were performed using Common Astronomy Software Application (CASA) version 5.4.1, along with the ALMA data analysis pipeline \citep{team}. During the calibration process, we employed standard pipeline tasks, including SETJY for flux density scaling of the flux calibrator, hifa\_bandpassflag for bandpass calibration, and hifa\_flagdata to remove problematic antenna data. Flux calibration was based on the Perley-Butler 2017 model \citep{per17}. Following the initial calibration, the science target G358.93--0.03 was extracted using the MSTRANSFORM task. The dust continuum emission images were generated from line-free channels using the TCLEAN task, employing the Hogbom deconvolution algorithm with a robust parameter of 0.5. Since dust continuum emission images and physical characteristics of dust continuum emissions have already been explored by \citet{manet23}, we do not revisit this study. Prior to spectral line imaging, continuum emission was subtracted from the UV data using the UVCONTSUB task. Subsequently, spectral images were created using the TCLEAN task with the SPECMODE = CUBE parameter across full frequency coverage. To enhance image quality and reduce noise levels, we performed iterative self-calibration using the GAINCAL and APPLYCAL tasks, including the parameter of solint = `inf'. Finally, primary beam correction was applied to both continuum and spectral images using the IMPBCOR task.

\section{Results}
\label{res}
\subsection{Spectral line emission towards G358.93--0.03}
The molecular spectra were extracted from both G358.93--0.03 MM1 and G358.93--0.03 MM3 using circular apertures with a diameter of 0.9$^{\prime\prime}$, which encompassed the regions exhibiting line emissions in both sources. These spectra are presented in Figure~2 of \citet{manet23}. Among the two sources, G358.93--0.03 MM1 displays a significantly richer molecular inventory than G358.93--0.03 MM3. In addition, its spectrum revealed an inverse P-Cygni profile in the emission lines of \ce{CH3OH}, a characteristic feature commonly associated with inward gas motion, suggesting an ongoing infall in G358.93--0.03 MM1. The synthesized beam sizes of the spectral images corresponding to the frequency ranges of 290.51--292.39 GHz, 292.49--294.37 GHz, 302.62--304.49 GHz, and 304.14--306.01 GHz are 0.42$^{\prime\prime}\times$0.36$^{\prime\prime}$, 0.42$^{\prime\prime}\times$0.37$^{\prime\prime}$, 0.41$^{\prime\prime}\times$0.36$^{\prime\prime}$, and 0.41$^{\prime\prime}\times$0.35$^{\prime\prime}$, respectively. The coordinates of G358.93--0.03 MM1 is RA (J2000) = 17$^{\mathrm{h}}$43$^{\mathrm{m}}$10$^{\mathrm{s}}$.101 and Dec (J2000) = --29$^\circ$51$^{\prime}$45$^{\prime\prime}$.693, while G358.93--0.03 MM3 is located at RA (J2000) = 17$^{\mathrm{h}}$43$^{\mathrm{m}}$10$^{\mathrm{s}}$.0144 and Dec (J2000) = --29$^\circ$51$^{\prime}$46$^{\prime\prime}$.193. The systemic velocities ($V_{LSR}$) of G358.93--0.03 MM1 and G358.93--0.03 MM3 are measured to be --16.5 km s$^{-1}$ and --18.2 km s$^{-1}$, respectively \citep{bro19}.
		
\subsubsection{Identification of c-\ce{C2H4O} and \ce{CH3CHO} towards G358.93--0.03 MM1}
We performed line identification and spectral line analysis using CASSIS \citep{vas15} with incorporation of the Cologne Database for Molecular Spectroscopy (CDMS; \citealt{mu05}) and the Jet Population Laboratory (JPL; \citealt{pic98}) molecular databases. After analysis of the spectral lines using both CDMS and JPL databases, we detected four rotational emission lines of c-\ce{C2H4O} and \ce{CH3CHO} towards G358.93--0.03 MM1 with a statistical significance of at least 3$\sigma$ across the observed frequency bands. After that, we fitted the local thermodynamic equilibrium (LTE) modelled spectra of both c-\ce{C2H4O} and \ce{CH3CHO} over the observed spectra. The LTE spectral model fitting was conducted using the Markov Chain Monte Carlo (MCMC) algorithm implemented within CASSIS. The LTE condition is reasonable for the inner region of G358.93--0.03 MM1 because the gas density of this hot core is $1\times10^{7}$ cm$^{-3}$ \citep{bro19}. The LTE-modelled fitted spectral line parameters are shown in Table~\ref{tab:MOLECULAR DATA}. 
		
To ensure accurate identification of non-blended transitions, we included over 200 molecular lines in our LTE model, incorporating transitions from species previously detected by \citet{manet23}, \citet{manet24}, \citet{man24a}, and \citet{man24d}. We found that all the detected emission lines of c-\ce{C2H4O} and \ce{CH3CHO} are non-blended. The upper-state energies ($E_{up}$) of the detected transitions of c-\ce{C2H4O} and \ce{CH3CHO} are found to be in the ranges of 59.44--70.08 K and 115.67--131.66 K, respectively. Importantly, no high-intensity transitions of c-\ce{C2H4O} and \ce{CH3CHO} are missing within the observed frequency ranges. The LTE-fitted rotational emission lines and corner diagrams are shown in Figures~\ref{fig:spectra} and \ref{fig:corner}. From the LTE model, we have derived that the column density and excitation temperature of c-\ce{C2H4O} are $(6.5\pm0.4)\times10^{15}$ cm$^{-2}$ and $180\pm22$ K, respectively. Similarly, the column density and excitation temperature of \ce{CH3CHO} are ($2.2\pm0.3)\times10^{16}$ cm$^{-2}$ and $160\pm25$ K, respectively. During LTE modelling, we used a source size of 0.42$^{\prime\prime}$.
		
\begin{table}
\centering
%\scriptsize 
\caption{Emitting regions of c-\ce{C2H4O} and \ce{CH3CHO} towards G358.93--0.03 MM1.}
\begin{adjustbox}{width=0.48\textwidth}
\begin{tabular}{cccccccc}
\hline 
Molecule&Rest frequency&Transition&$E_{u}$ &Emitting region\\
		&(GHz)         &          & (K)    &($\prime\prime$) \\
\hline
c-\ce{C2H4O}&291.431&8(3,5)--7(4,4)&70.08&0.42\\ 
			&291.478&10(0,10)--9(1,9)&79.01&0.40\\
			&291.882&5(4,2)--4(1,3)&32.53&0.41\\
			&305.033&7(5,3)--6(4,2)&59.44&0.41\\		
\hline
\ce{CH3CHO}&290.644&15(3,12)--14(3,11)$E$&131.63&0.41 \\
			&290.717&15(3,12)--14(3,11)$A$&131.66&0.40\\
			&293.694&15(1,14)--14(1,13)$E$&115.72&0.42\\
			&293.762&15(1,14)--14(1,13)$A$&115.67&0.41\\
\hline
\end{tabular}	
\end{adjustbox}
\label{tab:emitting region}
\end{table}
		
\begin{figure*}
\centering
\includegraphics[width=0.95\textwidth]{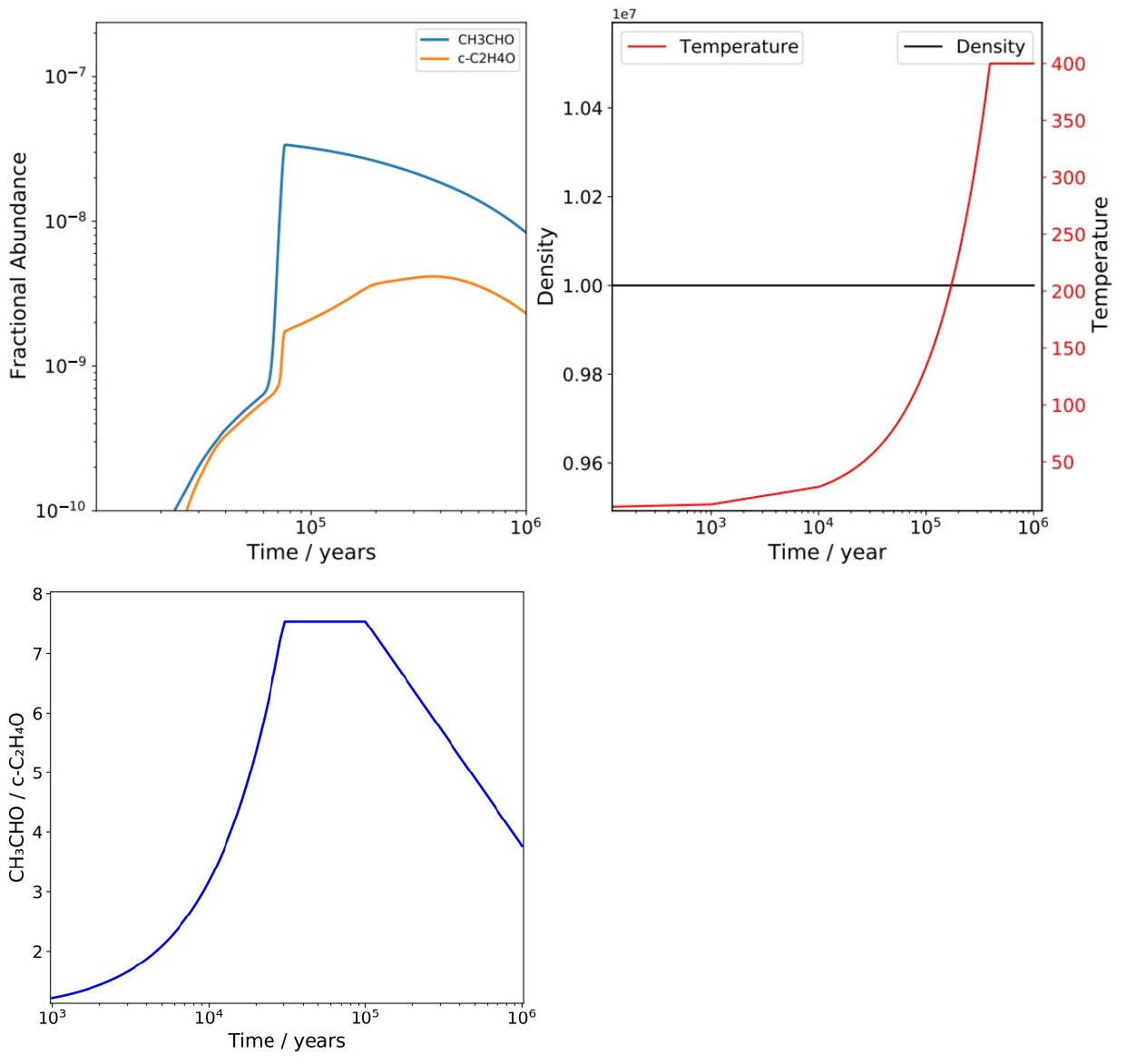}
\caption{Time-dependent gas-phase fractional abundances of c-\ce{C2H4O} and \ce{CH3CHO} (left panel) are computed using UCLCHEM over a timescale of 1$\times$10$^{5}$ years. The right panel displays the temperature profile during the warm-up phase. The bottom panel shows the time-dependent \ce{CH3CHO}/c-\ce{C2H4O} abundance ratio.}
\label{fig:model}
\end{figure*}
		
\begin{table}
\centering
%\scriptsize 
\caption{Initial fractional abundances of \ce{H2} and atomic elements at the start of the collapse phase.}
\begin{adjustbox}{width=0.27\textwidth}
\begin{tabular}{ccc}
\hline 
Species&  & Abundance\\
\hline
\ce{H2}&&4.99$\times$10$^{-1}$ \\
H & &2.00$\times$10$^{-3}$ \\
He & &9.00$\times$10$^{-2}$ \\
N & &7.50$\times$10$^{-5}$ \\
C & & 1.40$\times$10$^{-4}$ \\
O & & 3.20$\times$10$^{-4}$ \\
S & &8.00$\times$10$^{-8}$\\
Mg & &7.00$\times$10$^{-9}$\\
Na & &2.00$\times$10$^{-8}$\\
P & &3.00$\times$10$^{-9}$\\
Si & &8.00$\times$10$^{-9}$\\
Fe & &3.00$\times$10$^{-9}$\\
Cl & &4.00$\times$10$^{-9}$\\
\hline
\end{tabular}	
\end{adjustbox}
\label{tab:mod}
\end{table}
		
\begin{figure*}
\centering
\includegraphics[width=0.93\textwidth]{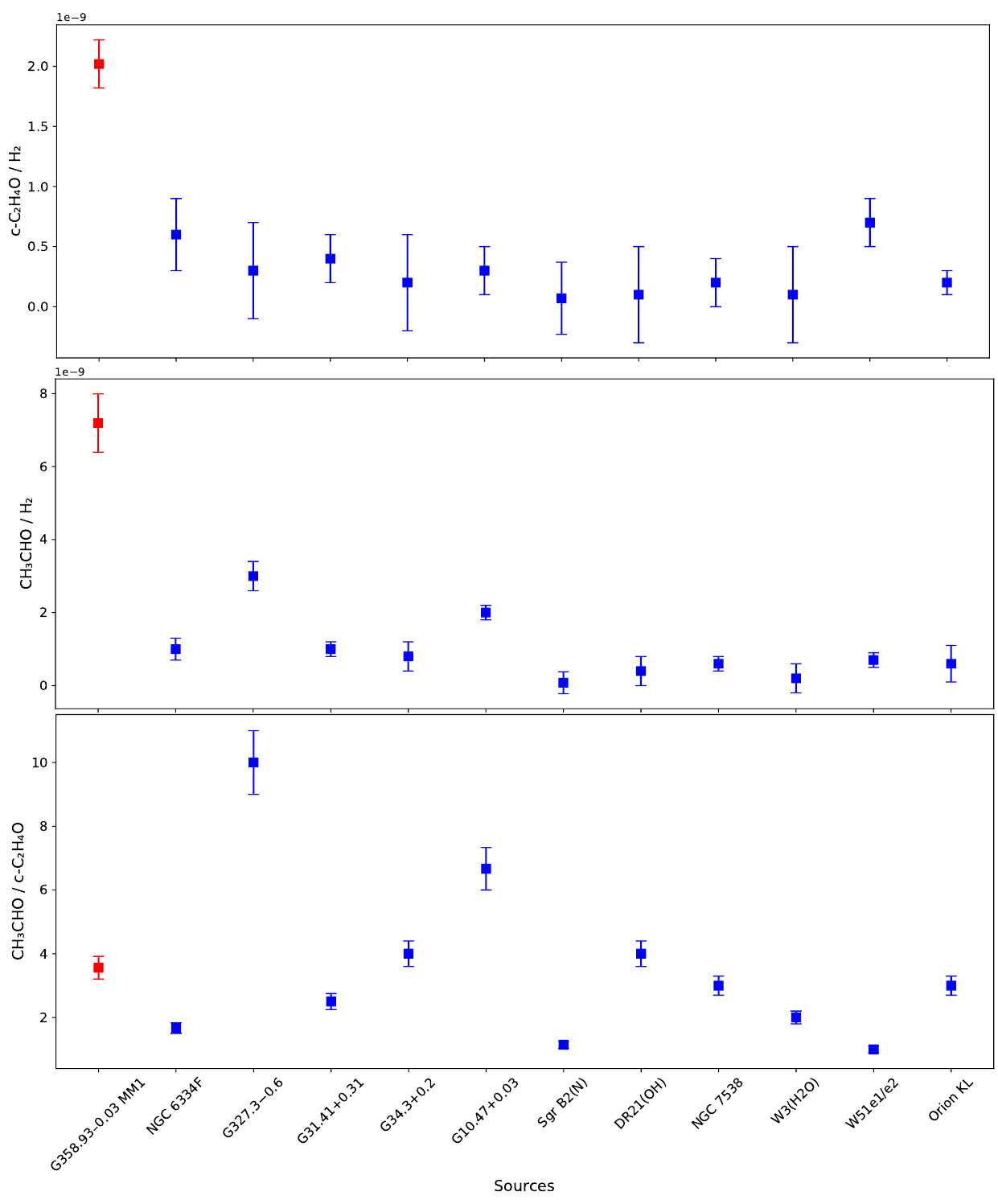}
\caption{Scatter plots showing the abundances of c-\ce{C2H4O} (top panel) and \ce{CH3CHO} (middle panel) relative to \ce{H2}, as well as the \ce{CH3CHO}/c-\ce{C2H4O} abundance ratio (bottom panel), towards G358.93--0.03 MM1 and other hot cores. In the plots, the red boxes with error bars represent the data points for G358.93--0.03 MM1, while the blue boxes with error bars represent the data points for other hot cores.}
\label{fig:bar}
\end{figure*}
		
\subsubsection{Abundances of c-\ce{C2H4O} and \ce{CH3CHO} towards G358.93--0.03 MM1}
The fractional abundances of c-\ce{C2H4O} and \ce{CH3CHO} were estimated by dividing the column density, measured within 0.50$^{\prime\prime}$, by the column density of molecular hydrogen (\ce{H2}). The fractional abundances of c-\ce{C2H4O} and \ce{CH3CHO} relative to \ce{H2} are $(2.1\pm0.2)\times10^{-9}$ and $(7.1\pm0.9)\times 10^{-9}$, where the column density of \ce{H2} towards G358.93--0.03 MM1 is $(3.1\pm0.2)\times10^{24}$ cm$^{-2}$ \citep{manet23}. The value of the column density of hydrogen is derived from the dust continuum emission at a wavelength of 988 $\mu$m \citep{manet23}. The column density ratio between \ce{CH3CHO} and c-\ce{C2H4O} is $3.4\pm0.7$. Previously, \citet{ik01} reported that \ce{CH3CHO}/c-\ce{C2H4O} abundance ratio in hot cores ranged from 1 to 3. Our derived value toward G358.93--0.03 MM1 falls within this range, consistent with their findings. This result suggests that the physical and chemical properties of \ce{CH3CHO} and c-\ce{C2H4O} in G358.93--0.03 MM1 are similar to those observed in the hot cores studied by \citet{ik01}.
		
\subsubsection{Rotational diagram of c-\ce{C2H4O} and \ce{CH3CHO}}
To determine the rotational temperature ($T_{rot}$) and total column density ($N_{T}$) of c-\ce{C2H4O} and \ce{CH3CHO}, we constructed a rotational diagram based on non-blended transitions. This method serves to validate the estimated excitation temperature and column density obtained from LTE spectral modelling. We assumed that the observed emission lines of c-\ce{C2H4O} and \ce{CH3CHO} are optically thin and that the population distribution follows LTE. For optically thin molecular lines, the column density is calculated using the following expression \citep{gol99}:
			
\begin{equation}
{N_u^{thin}}=\frac{3{g_u}k_B}{8\pi^{3}\nu S\mu^{2}}\int{T_{mb}dV}
\label{eq:rotd1}
\end{equation}
In Equation~\ref{eq:rotd1}, $g_u$ denotes the upper-state degeneracy, $k_B$ is the Boltzmann constant, $\mu$ is the electric dipole moment of the molecule, $\nu$ represents the rest frequency, $S$ is the line strength, and $\int T_{mb},dV$ corresponds to the integrated line intensity in units of km s$^{-1}$. Under LTE conditions, the total column density of the molecule is given by:
			
\begin{equation}
\frac{N_u^{thin}}{g_u} = \frac{N_{total}}{Q(T_{rot})}\exp\left(\frac{-E_u}{k_BT_{rot}}\right)
\label{eq:rotd2}
\end{equation}
In Equation~\ref{eq:rotd2}, $T_{\mathrm{rot}}$ denotes the rotational temperature of the molecule, and $Q(T_{\mathrm{rot}})$ is the partition function evaluated at that temperature. For c-\ce{C2H4O}, the values of $Q(T_{\mathrm{rot}})$ are 39396.23 at 300 K, 25584.66 at 225 K, 13927.78 at 150 K, and 4929.48 at 75 K taken from CDMS \citep{mu05}. Similarly, the values of $Q(T_{\mathrm{rot}})$ for \ce{CH3CHO} are 86841.08 at 300 K, 50049.72 at 225 K, 22892.20 at 150 K, and 6495.80 at 75 K, taken from the JPL database \citep{pic98}. By rearranging Equation~\ref{eq:rotd2}, it can be expressed as:
			
\begin{equation}
\ln\left(\frac{N_u^{thin}}{g_u}\right) = \ln(N)-\ln(Q)-\left(\frac{E_u}{k_BT_{rot}}\right)
\label{eq:rotd3}
\end{equation}
Equation~\ref{eq:rotd3} demonstrates a linear relationship between $E_u$ and $\ln(N_u/g_u)$. The values of $\ln(N_u/g_u)$ were derived using Equation~\ref{eq:rotd2}. According to Equation~\ref{eq:rotd3}, the spectral parameters of various transitions of c-\ce{C2H4O} and \ce{CH3CHO} should be fitted with a straight line, the slope of which is inversely proportional to $T_{\mathrm{rot}}$ and the value of $N_{T}$ is estimated from its intercept. To construct the rotational diagram, we extracted the spectral line parameters by fitting a single Gaussian model to the non-blended lines of c-\ce{C2H4O} and \ce{CH3CHO}. The resulting rotational diagram is presented in Figures~\ref{fig:rotd}. From this analysis, we derive the $N_{T}$ and $T_{\mathrm{rot}}$ of c-\ce{C2H4O} to be $(5.2\pm0.3)\times10^{15}$ cm$^{-2}$ and $177\pm17$ K, respectively. Similarly, for \ce{CH3CHO}, we obtain the value of $N_{T}$ and $T_{\mathrm{rot}}$ are $(1.6\pm0.2)\times10^{16}$ cm$^{-2}$ and $158\pm32$ K. The estimated values of $N_{T}$ and $T_{\mathrm{rot}}$ for both c-\ce{C2H4O} and \ce{CH3CHO} derived from the rotational diagram analysis are in close agreement with those obtained from LTE spectral modelling. This consistency between $T_{\mathrm{rot}}$ and the excitation temperature indicates that both molecules are likely in LTE conditions within the hot core G358.93--0.03 MM1.
		
\begin{figure*}
\centering
\includegraphics[width=1.0\textwidth]{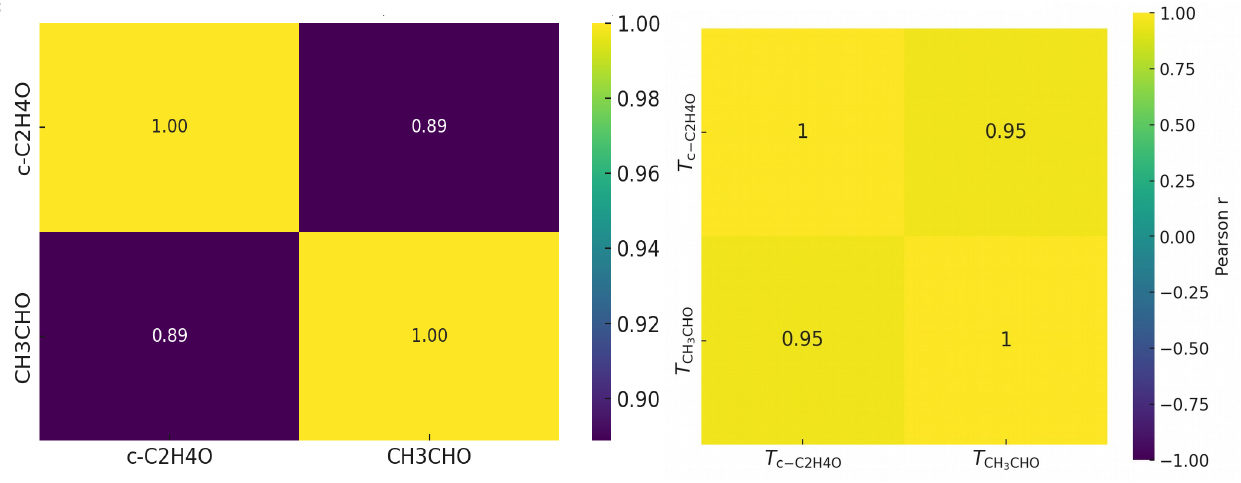}
\caption{Pearson correlation coefficient heat maps for abundances relative to \ce{H2} (left panel) and excitation temperatures (right panel) of c-\ce{C2H4O} and \ce{CH3CHO}. The colour indicates the strength of the correlation, with the corresponding Pearson coefficient ($r$) displayed in each cell.}
\label{fig:corr}
\end{figure*}

\subsubsection{Studies of c-\ce{C2H4O} and \ce{CH3CHO} towards G358.93--0.03 MM3}
We also investigated the rotational emission lines of both c-\ce{C2H4O} and \ce{CH3CHO} toward G358.93--0.03 MM3. However, no clear transitions were observed. Based on LTE spectral modelling, the estimated upper limit column densities for c-\ce{C2H4O} and \ce{CH3CHO} are $\leq(1.2\pm0.6)\times10^{13}$ cm$^{-2}$ and $\leq(8.2\pm0.52)\times10^{12}$ cm$^{-2}$, respectively. The upper limits for the fractional abundances are $\leq(3.6\pm1.9)\times10^{-11}$ for c-\ce{C2H4O} and $\leq(2.4\pm0.5)\times10^{-11}$ for \ce{CH3CHO}, where the value of $N$(\ce{H2}) towards G358.93--0.03 MM3 is $(3.5\pm0.7)\times10^{23}$ cm$^{-2}$ \citep{manet23}.

\subsubsection{Searching of vinyl alcohol towards G358.93--0.03 MM1}
Using the LTE-modelled spectra, we also searched for the emission lines of vinyl alcohol (\ce{CH2CHOH}), the third isomer of c-\ce{C2H4O}, and \ce{CH3CHO}. Unfortunately, we did not detect these molecules in the spectra of G358.93--0.03 MM1. Assuming an excitation temperature of 170 K, we estimate the upper limits for the column densities of $\leq(2.35\pm0.36)\times10^{12}$ cm$^{-2}$ for the $syn$-conformer and $\leq(4.46\pm0.58)\times10^{11}$ cm$^{-2}$ for the $anti$-conformer. Despite the high sensitivity of our spectra, these limits are not particularly stringent because the transitions of \ce{CH2CHOH} in the observed frequency range are inherently weak, in contrast to c-\ce{C2H4O} and \ce{CH3CHO}. As a result, we cannot draw definitive conclusions about the relative abundance of \ce{CH2CHOH} compared to that of c-\ce{C2H4O} and \ce{CH3CHO}.
		
\begin{table*}
\centering
%\scriptsize 
\caption{Peak gas-phase fractional abundances from the warm-up model and their comparison with observed values.}
\begin{adjustbox}{width=1.0\textwidth}
\begin{tabular}{|c|cc|cc|c|c}
\hline 
Molecule & Modelled abundance & Modelled temperature& Observed abundance & Excitation temperature & \multicolumn{1}{c|}{Dominant reactions} \\
		&                  &  (K)                 &                  &           (K)             &      \\
\hline
c-\ce{C2H4O}/\ce{H2} &$2.85\times10^{-9}$ &210 &$(2.1\pm0.2)\times10^{-9}$&$180\pm22$&O + \ce{C2H4} $\longrightarrow$ c-\ce{C2H4O} \\
\hline
\ce{CH3CHO}/\ce{H2}  &$2.15\times10^{-8}$  &185 &$(7.1\pm0.9)\times10^{-9}$&$160\pm25$& \ce{CH3} + HCO $\longrightarrow$ \ce{CH3CHO} \\
\hline
\ce{CH3CHO}/c-\ce{C2H4O}&7.54&-- &$3.4\pm0.7$&-- &-- \\
\hline
\end{tabular}	
\end{adjustbox} 
\label{tab:abun}
\end{table*}
		
\subsection{Spatial distribution of c-\ce{C2H4O} and \ce{CH3CHO}}
We generated integrated emission maps (moment zero maps) of the detected rotational emission lines of c-\ce{C2H4O} and \ce{CH3CHO} using the IMMOMENTS task. During this process, we selected the channel ranges from the spectral images where the c-\ce{C2H4O} and \ce{CH3CHO} emission lines are present. These moment zero maps for the non-blended transitions of c-\ce{C2H4O} and \ce{CH3CHO} towards G358.93--0.03 MM1 are displayed in Figure~\ref{fig:intmap}. In addition, we overlay the 988 $\mu$m continuum emission map of G358.93--0.03, taken from \citet{manet23}, onto the integrated emission maps. Our analysis revealed that the emission peaks of c-\ce{C2H4O} and \ce{CH3CHO} coincide with the dust continuum emission. This suggests that the rotational emission lines of c-\ce{C2H4O} and \ce{CH3CHO} originate from the dense, warm inner regions of G358.93--0.03 MM1. Following this, we applied the CASA task IMFIT to fit a 2D Gaussian to the integrated emission maps and estimate the size of the emitting regions. The results for the derived emitting regions at different frequencies are summarized in Table~\ref{tab:emitting region}. The synthesized beam sizes for the integrated emission maps of both molecules ranged from 0.41$^{\prime\prime} \times $0.36$^{\prime\prime}$ to 0.42$^{\prime\prime}\times$0.37$^{\prime\prime}$, whereas the derived emitting region sizes for c-\ce{C2H4O} and \ce{CH3CHO} varied between 0.40$^{\prime\prime}$ and 0.42$^{\prime\prime}$. We observed that the derived emitting regions of both molecules are consistent with the source size used in the LTE modelling. This consistency indicates that the physical parameters we derived, such as column density and excitation temperature, are appropriate for both molecules. After fitting a 2D Gaussian, we observed that the sizes of the emitting regions are comparable to or slightly larger than the synthesized beam, implying that the detected emission lines of c-\ce{C2H4O} and \ce{CH3CHO} are either unresolved or marginally resolved toward G358.93--0.03 MM1. Since the integrated intensity images of both molecules are unresolved or marginally resolved, any conclusions regarding the spatial morphology of both molecules toward G358.93--0.03 MM1 are limited. Observations with higher spatial and angular resolutions are necessary to determine the spatial distributions of c-\ce{C2H4O} and \ce{CH3CHO} in this region accurately. 
		
\subsection{Chemical model of c-\ce{C2H4O} and \ce{CH3CHO}}
We computed out a two-phase (gas + grain) warm-up chemical model for c-\ce{C2H4O} and \ce{CH3CHO} using the time-dependent gas-grain chemical code UCLCHEM \citep{hold17} with the aim of exploring their modelled abundances and possible formation pathways.  This chemical code incorporates both thermal and non-thermal desorption processes in the gas-phase and grain surface chemistry, considering various physical conditions and astrophysical environments within the ISM. This chemical code solves reaction rate equations to estimate the abundance of gas-phase and grain surface molecules relative to hydrogen across various environments where these molecules are present \citep{viti13}. To simulate the physical conditions in the hot core, we utilized a two-phase warm-up chemical model comprising an initial free-fall collapse, followed by a static warm-up phase. In the first stage (Phase I), representing the cold collapse, the gas density ($n_{\mathrm{H}_2}$) increases from $1\times10^{2}$ cm$^{-3}$ to $1\times10^{7}$ cm$^{-3}$, with the temperature of the gas and the dust constant at 8 K. For chemical modelling, we adopted a cosmic ray ionization rate ($\zeta$) of $1.3\times10^{-17}$ s$^{-1}$ \citep{gar08}. According to the density profile derived from observations and theoretical studies of hot core regions by \citet{no04}, the gas density ($n_{\mathrm{H}_2}$)  remains relatively uniform within a radius of 0.1 pc. In chemical model, the initial visual extinction ($A_{\mathrm{V0}}$) can be estimated as follows:
				
\begin{equation}
A_{\mathrm{V}} = A_{\mathrm{V0}} \left( \frac{n_{\mathrm{H_{2}}}}{n_{\mathrm{H0}}} \right)^{2.3}
\label{eq:visual}
\end{equation}
In Equation~\ref{eq:visual}, $n_{\mathrm{H0}}$ represents the initial gas density (1 $\times$ 10$^{2}$ cm$^{-3}$) and $A_{\mathrm{V0}}$ is taken as 2 \citep{gar08}, as defined at the onset of the quasistatic collapse phase. During this phase, the accretion rate of atoms and molecules on the grain surfaces is 10$^{-5}$ \textup{M}$_{\odot}$ yr$^{-1}$, and it varies depending on the gas density within the hot cores \citep{vi04}. We assumed a sticking probability of unity during chemical modelling, implying that all incoming hydrogen atoms will stick to the grain surface upon encountering an inactive site. During this phase, molecular species can become hydrogenated or react quickly with other species on the grain surface. The chemical model used the initial atomic abundances relative to the solar values, as shown in Table~\ref{tab:mod}. These values were adopted from \citet{gar13} and represent the conditions at the beginning of the collapse phase. 
			
In Phase II (the warm-up stage), the gas temperature increased from 8 K to 400 K, whereas the hydrogen density ($n_{\mathrm{H}_2}$) remained constant at $1\times10^{7}$ cm$^{-3}$. The temperature evolution during this stage follows the expression proposed by \citet{gar13}:  
			
\begin{equation} 
T = T_{0} + (T_{max}-T_{0})(\Delta t/t_h)^n 
\label{tab:temp} 
\end{equation} 
In Equation~\ref{tab:temp}, $T_0$ = 8 K represents the starting temperature at the onset of the warm-up phase, and $T_{\mathrm{max}}$ is the peak temperature reached by the end of this phase. Parameter $\Delta t$ defines the total duration of the warm-up phase, and $t_h$ is its characteristic timescale. This stage is followed by a first-order process, with $n = 1$. In our study, we adopted the fiducial values of $1\times10{^5}$ yr and $T_{max}$ = 400 K. Previous chemical models by \cite{gar08}, \cite{gar13}, and \cite{suz18} showed that a time scale of $1\times10{^5}$ yr is appropriate for the evolution of complex molecules during the warm-up phase. It is generally assumed that the gas and dust temperatures remain thermally coupled during this stage. Given that coupling and the dust temperature of G358.93--0.03 MM1 being 150 K, a maximum temperature of 400 K is reasonable in our chemical modelling. In this stage, the molecules stop freezing on grain surfaces, and both thermal and non-thermal desorption mechanisms contribute to the release of surface-bound species back into the gas phase. Additionally, our model includes volcanic desorption, monomolecular desorption, grain mantle sublimation, and co-desorption with \ce{H2O}, as described in Tables 1 and 2 in \citet{vi04}. The two-phase warm-up chemical model employed here follows an approach similar to that used in \citet{gar08}, \citet{occ14}, \citet{man24c} and \citet{man24e}.
			
In our chemical modelling, we incorporated over 100 reactions (both gas-phase and grain surface) involving c-\ce{C2H4O} and \ce{CH3CHO}, sourced from \citet{gar08}, \cite{occ14}, and \citet{gar13}, as well as the astrochemical databases KIDA \citep{wak} and UMIST \citep{mc13}. Most of the chemical reactions utilized in this work were taken from UMIST and KIDA databases, which provide comprehensive and validated reaction networks. These reactions were incorporated into a two-phase warm-up chemical model, designed to simulate the progressive thermal evolution characteristic of hot cores in star-forming regions. Accordingly, the chemical network employed in this study is considered sufficiently complete for tracing the formation and destruction pathways of the complex organic molecules under investigation. The binding energies (E$_D$) adopted for c-\ce{C2H4O} and \ce{CH3CHO} were 2450 K and 5400 K, respectively, based on the values reported by \cite{occ14} and \citet{suz18}. Figure~\ref{fig:model} shows the time evolution of the fractional abundances (relative to hydrogen) of c-\ce{C2H4O} and \ce{CH3CHO} over a period of $1\times10^{5}$ yr. We also computed the time-dependent \ce{CH3CHO}/c-\ce{C2H4O} ratio profile in the lower panel of Figure~\ref{fig:model}. The corresponding peak gas-phase abundances and associated temperatures for c-\ce{C2H4O} and \ce{CH3CHO} are summarized in Table~\ref{tab:abun}. The gas-phase formation of c-\ce{C2H4O} was relatively inefficient. Instead, these molecules are predominantly produced on the surfaces of the dust grains through the reaction between O and \ce{C2H4}, which is released during the warm-up stage in the gas-phase. The key reaction of the formation of c-\ce{C2H4O} on grain surfaces is:\\\\ 
O + \ce{C2H4} $\longrightarrow$ c-\ce{C2H4O}~~~~~~~~~~(3)\\\\ 
Similarly, in the warm-up stage, \ce{CH3CHO} is formed via a reaction between \ce{CH3} and HCO, which is released in the gas-phase. The formation reaction of \ce{CH3CHO} is:\\\\  
\ce{CH3} + HCO $\longrightarrow$ \ce{CH3CHO}~~~~~~~~(4)\\\\ 
The formation of molecules on the grain surfaces is primarily driven by the mobility of \ce{CH3}, as it diffuses more readily than HCO, which faces a higher energy barrier. Acetaldehyde reaches its peak surface abundance at approximately 30 K, and this abundance remained largely unchanged until thermal desorption became prominent near 50 K. After simulation, reactions 3 and 4 are found to be the dominant reactions for the formation of c-\ce{C2H4O} and \ce{CH3CHO} on the grain surface.
			
\section{Discussion}
\label{dis}
\subsection{Abundance comparison and correlation studies}
We performed a comparative analysis of the fractional abundances of c-\ce{C2H4O} and \ce{CH3CHO} (relative to \ce{H2}) in G358.93--0.03 MM1 and selected well-known hot cores. In addition to individual abundances, we examined the abundance ratio \ce{CH3CHO}/c-\ce{C2H4O} to explore the potential chemical trends. The comparison included sources such as NGC 6334F, G327.3--0.6, G31.41+0.31, G34.3+0.2, G10.47+0.03, Sgr B2(N), DR21(OH), NGC 7538, W3(H2O), W51 e1/e2, and Orion KL \citep{ik01}. As illustrated in Figure~\ref{fig:bar}, the scatter plots show the relative abundances of c-\ce{C2H4O} and \ce{CH3CHO} for these sources.
			
Scatter plot analysis reveals that G358.93--0.03 MM1 exhibits not only one of the highest absolute abundances of both c-\ce{C2H4O} and \ce{CH3CHO}, but also a moderate \ce{CH3CHO}/c-\ce{C2H4O} ratio (3.4). This contrasts with sources like G327.3--0.6, where the ratio is much higher (10), implying a disproportionate \ce{CH3CHO} formation or stronger depletion of c-\ce{C2H4O}. In sources such as W51 e1/e2 and G10.47+0.03, the near-unity ratios suggest co-evolution or closely linked formation mechanisms. These variations in the \ce{CH3CHO}/c-\ce{C2H4O} ratio across different sources may reflect differences in physical conditions, such as temperature, UV flux, and warm-up timescales, which affect grain surface chemistry and desorption rates. Additionally, the elevated abundance of c-\ce{C2H4O} in G358.93--0.03 MM1, coupled with a moderate \ce{CH3CHO}/c-\ce{C2H4O} ratio, indicates a chemically rich and possibly less-evolved hot core, where both molecules are efficiently formed and preserved. Furthermore, the variation in the \ce{CH3CHO}/c-\ce{C2H4O} ratios across different sources suggests diversity in the chemical pathways and evolutionary stages among the hot cores. These findings imply that the efficiency of the formation and destruction mechanisms for these O-bearing species may differ substantially across regions, potentially influenced by factors such as temperature, grain surface chemistry, and warm-up timescales. This comparative molecular study underscores the importance of observing multiple related species to trace the chemical pathways and evolutionary status in high-mass star-forming regions.
			
Figure~\ref{fig:corr} displays the Pearson correlation matrix heat map illustrating the relationships between the abundances and excitation temperatures of c-\ce{C2H4O} and \ce{CH3CHO}. We found that the abundances of c-\ce{C2H4O} and \ce{CH3CHO} are strongly correlated ($r = 0.89$), which indicates that both molecules may be chemically linked. The excitation temperatures of c-\ce{C2H4O} and \ce{CH3CHO} exhibited a strong and statistically significant positive correlation ($r$ = 0.95) across multiple hot cores, including G358.93--0.03 MM1. This suggests that these molecules likely coexist in similar thermal environments and may share the related formation or excitation mechanisms. The consistent trend across diverse sources supports the idea of a physical or chemical link between these two species in hot core chemistry.
			
\subsection{Comparison between observed and modelled abundances}
To understand the formation pathways of c-\ce{C2H4O} and \ce{CH3CHO} towards G358.93--0.03 MM1, we compared their observed abundance with the abundance simulated by two-phase warm-up chemical modelling using UCLCHEM. This comparison is physically justified, as the density of G358.93--0.03 MM1 ($1 \times 10^{7}$ cm$^{-3}$, \citealt{bro19}) is similar to the gas density adopted in the chemical model. The comparison of the observed and modelled abundances is listed in Table~\ref{tab:abun}. To investigate the pathways of c-\ce{C2H4O} and \ce{CH3CHO} formation in G358.93--0.03 MM1, we found that the observed abundance of c-\ce{C2H4O} relative to \ce{H2} is similar to the modelled value within a factor of 0.73. This indicates that c-\ce{C2H4O} might be formed towards G358.93--0.03 MM1 through the reaction between \ce{C2H4} and O on the grain surface (reaction 3). We also found that the modelled abundance of \ce{CH3CHO} is approximately one order of magnitude higher than the observed value. This indicates that our chemical model does not explain the formation pathways of \ce{CH3CHO} towards G358.93--0.03 MM1. The modelled \ce{CH3CHO}/c-\ce{C2H4O} ratio is higher than the observed value. To understand the formation pathway of \ce{CH3CHO} towards G358.93--0.03 MM1, we compared our observed abundance with the modelled abundance of \cite{gar08}. After chemical modelling, \cite{gar08} derived that modelled abundances of \ce{CH3CHO} in the context of hot cores are $2.5 \times 10^{-9}$, $2.2 \times 10^{-8}$, and $9.7 \times 10^{-9}$, corresponding to the first, medium, and slow warm-up conditions, respectively. We found that our modelled abundance of \ce{CH3CHO} using UCLCHEM is similar to the medium warm-up modelled value of \cite{gar08}. We note that our observed abundance of \ce{CH3CHO} is nearly identical within a factor of 0.74 to the slow warm-up modelled value of \cite{gar08}. During a chemical simulation, \cite{gar08} considered that reaction 4 is the dominant pathway for the production of \ce{CH3CHO}. Since the observed and modelled abundances are close, that indicates \ce{CH3CHO} might be formed via the grain surface reaction between \ce{CH3} and HCO (reaction 4) towards G358.93--0.03 MM1.
			
\section{Summary and conclusions}
\label{con}
In this article, we present the first detection of rotational emission lines of c-\ce{C2H4O} and \ce{CH3CHO} towards G358.93--0.03 MM1. The column density and excitation temperature of c-\ce{C2H4O}, derived from the LTE spectral modelling, are $(6.5 \pm 0.4)\times 10^{15}$ cm$^{-2}$ and 180 $\pm$ 22 K, respectively. For \ce{CH3CHO}, the column density and excitation temperature are (2.2 $\pm$ 0.3)$\times$10$^{16}$ cm$^{-2}$ and 160 $\pm$ 25 K, respectively. The fractional abundances of c-\ce{C2H4O} and \ce{CH3CHO} relative to \ce{H2} are $(2.1 \pm 0.2) \times 10^{-9}$ and $(7.1 \pm 0.9) \times 10^{-9}$, respectively. The column density ratio between \ce{CH3CHO} and c-\ce{C2H4O} is 3.4 $ \pm$ 0.72. The Pearson correlation heat map revealed strong positive correlations between the abundance and excitation temperatures of c-\ce{C2H4O} and \ce{CH3CHO}, suggesting that these two molecules may be chemically linked. To explore their potential formation pathways, we developed a two-phase warm-up chemical model using the gas-grain chemical code UCLCHEM. We propose that c-\ce{C2H4O} may be formed in G358.93--0.03 MM1 via a grain surface reaction between \ce{C2H4} and O. Similarly, \ce{CH3CHO} may be produced through the grain surface reaction between \ce{CH3} and HCO. These findings enhance our understanding of the complex oxygen chemistry occurring in the star-forming regions and the formation pathways of organic molecules in the ISM.
			
\section*{ACKNOWLEDGEMENTS}
We thank the anonymous reviewer for his/her constructive comments, which have significantly improved the quality of the manuscript. This paper makes use of the following ALMA data: ADS /JAO.ALMA\#2019.1.00768.S. ALMA is a partnership of ESO (representing its member states), NSF (USA), and NINS (Japan), together with NRC (Canada), MOST and ASIAA (Taiwan), and KASI (Republic of Korea), in co-operation with the Republic of Chile. The Joint ALMA Observatory is operated by ESO, NRAO, and NAOJ.
			
\section*{DATA AVAILABILITY}
The plots within this paper and other findings of this study are available from the corresponding author on reasonable request. The data used in this paper are available in the ALMA Science Archive (\url{https://almascience.nrao.edu/asax/}), under project code 2019.1.00768.S.
			
\section*{Conflicts of interest}{The authors declare no conflict of interest.}
			
\section*{Credit authorship contribution statement}
Arijit Manna: analyzed the ALMA data, analyzed the spectral lines, and wrote the original draft. Sabyasachi Pal: Conceptualized the project and edited the draft.

\end{document}